\documentstyle[12pt]{article}
\begin {document}
\newcommand {\beq}{\begin{equation}}
\newcommand {\beqa}{\begin{eqnarray}}
\newcommand {\eeq}{\end{equation}}
\newcommand {\eeqa}{\end{eqnarray}}
\newcommand {\n}{\nonumber \\}
%%%%%%%%%%%%%%%%%%%%%%%%%%%%%%%%%%%%%%%%%%%%%%%%%%%%%%%%%%%%%%%%%%%%%%%%%
\begin{flushright}
UT-599
\\
March  1992
\end{flushright}
\begin{center}
{\large\bf BRST-Fixed Points  \\
      and \\
      Topological Conformal Symmetry} \vspace{.7 in}\\
{\bf Toshio Nakatsu and
        Yuji Sugawara \vspace{.5 in}\\
        {\it Department of Physics, University of Tokyo}\\
        {\it Bunkyo-ku,Tokyo 113,Japan \vspace{.8 in}}}
\end{center}
\begin{abstract}
  We study the twisted version of the supersymmetric
$G/T=SU(n)/U(1)^{\otimes (n-1)}$ gauged Wess-Zumino-Witten
model.
By studying its fixed points under
BRST transformation
this model is shown to be reduced to a simple
topological field theory,
that is, the topological matter system in the K.Li's
theory of 2 dimensional gravity for the case of
$n=2$, and its generalization for $n \geq 3$.
\end{abstract}

%%%%%%%%%%%%%%%%%%%%%%%%%%%%%%%%%%%%%%%%%%%%%%%%%%%%%%%%%%%%%%%%%%%%%%%%%%%%
\newpage

{}~

The idea of BRST-fixed points has been introduced into
the study of some topological field theories (TFTs) with
BRST symmetries \cite{Witten}.
It has been pointed out
\cite{Witten},\cite{Eguchi}
that the path-integration of BRST-invariant field
theories is reduced to an integral
over fixed point sets of BRST transformation
and the original TFT may be reduced to a
much simple one.
In this article we will show that the twisted version of
the supersymmetric $G/T=SU(n)/U(1)^{\otimes{(n-1)}}$
gauged Wess-Zumino-Witten (WZW) model
is reduced to a simple supersymmetric Coulomb gas system.
This system coincides with
the topological matter system in the K.Li's theory of
2 dimensional gravity for the case of $n=2$, and its generalization for
$n \geq 3$.
We argue that the reduction to fixed points in operator language
is given by the cancellation mechanism between bosonic
and fermionic ghosts known in the theory
of Hamiltonian reduction (quartet mechanism).

{}~

                 The topological conformal model \cite{EY}
which is obtainable by twisting the N=2 superconformal model
based on the flag manifold $G/T=SU(n)/U(1)^{\otimes (n-1)}$
\cite{KazamaSuzuki} was known to be described as two dimensional
Lagrangian field theory \cite{Witten}.
The action of them is given by;
\beqa
I(g,\chi,\psi,A)=
kS_{G}(g,A)
+S_{\chi \psi}(\chi,\bar{\chi},\psi,\bar{\psi},A)~~~.
\label{eq:action1}
\eeqa
The bosonic part of the action $S_{G}$ (\ref{eq:action1})
is that of the $G=SU(n)$ WZW model
gauged by the Cartan subgroup $T=U(1)^{\otimes (n-1)}$,
which can be obtained by adding the following gauge interaction
to $S_{G}^{WZW}(g)$, the action of an ungauged $G$ WZW model;
\beqa
&&kS_{G}(g,A) \n
&&=kS_{G}^{WZW}(g)
+\frac{i k}{2 \pi}\int_{\Sigma}d^2z
\mbox{Tr}
\{-g^{-1}\bar{\partial}g  A_{z}
+A_{\bar{z}} \partial gg^{-1}
-A_{\bar{z}} gA_{z}g^{-1}
+A_{\bar{z}} A_{z} \}~~. \n
&&~~~~~
\label{eq:action2}
\eeqa
where $g$ is a chiral field on a Riemann surface $\Sigma$,
taking its value in $G$, and
$A=A_{z}dz+A_{\bar{z}}d \bar{z}$
\footnote{$A_{z},A_{\bar{z}}$ satisfy $A_{z}^{\dag}=-A_{\bar{z}}$ .}
is $T=U(1)^{\otimes (n-1)}$ gauge field on $\Sigma$ .
The fermionic part of the action (\ref{eq:action1})
is ;
\beqa
S_{\chi \psi}(\chi,\bar{\chi},\psi,\bar{\psi},A)
= \frac{1}{2 \pi i}\int_{\Sigma}d^2z
\mbox{Tr}\{\chi_{z} \bar{\partial}_{A_{\bar{z}}}\psi
+\bar{\chi}_{\bar{z}} \partial_{A_{z}} \bar{\psi}\}~~~.
\label{eq:action3}
\eeqa
The fermionic fields
$\chi,\bar{\chi},\psi,\bar{\psi}$
are introduced in the following way:
The ghost fields $\psi,\bar{\psi}$
behave as $(0,0)$-forms on $\Sigma$,
taking their values in
$m_{+},m_{-}$ respectively.
Here $m_{\pm}$ are the components of the Cartan-Weyl decomposition
of $su(n)$ : $su(n)=t \oplus m_{+}
\oplus m_{-}$.
The anti-ghost fields
$\chi=\chi_{z}dz,\bar{\chi}=\bar{\chi}_{\bar{z}}d \bar{z}$
behave as
$(1,0),(0,1)$-forms on $\Sigma$, taking their values
in $m_{-},m_{+}$ respectively.
Note that the space $m_{+}\oplus m_{-}$ can be regarded as
the (complexified) tangent space of the flag manifold
$G/T=SU(n)/U(1)^{\otimes (n-1)}$.
Thus we will call these fermionic fields as coset ghosts.
The covariant derivative
$d_{A}=\partial_{A_{z}}+\bar{\partial}_{A_{\bar{z}}}$
is defined via the adjoint action of the Cartan subgroup $T$
on the flag manifold;$d_{A}=d+\mbox{[}A,\cdot~ \mbox{]}$.

                          The action (\ref{eq:action1}) has
the fermionic symmetry of the following type
\footnote{We parametrize the coset (anti-)ghosts such as
$\psi=\sum_{\alpha \in \Delta_{+}}\psi^{\alpha}e_{\alpha}
{}~(\chi_{z}=\sum_{\alpha \in \Delta_{+}}\chi_{z \alpha}e^{\alpha})$.
Here $\Delta_{+}$ is the set of the positive roots of $su(n)$,
and $\{e_{\alpha}\}_{\alpha \in \Delta_{+}}$,
$\{e^{\alpha}\}_{\alpha \in \Delta_{+}}$
are the Cartan-Weyl basis of $m_{+}$,$m_{-}$ respectively.}
;
\beqa
\delta_{G/T}g=\psi g~,
&&
\bar{\delta}_{G/T}g=-g \bar{\psi}~~, \n
\delta_{G/T}\chi_{z,\alpha}=\mbox{Tr}e_{\alpha}
(k\partial_{A_{z}}gg^{-1}+\mbox{[}\chi_{z},\psi \mbox{]})~,
&&
\bar{\delta}_{G/T}\bar{\chi}_{\bar{z}}^{\alpha}=
\mbox{Tr}e^{\alpha}(-kg^{-1}\bar{\partial}_{A_{\bar{z}}}g
+\mbox{[}\bar{\chi}_{\bar{z}},\bar{\psi}\mbox{]})~~,\n
\delta_{G/T}\psi =
\frac{1}{2}\mbox{[}\psi,\psi \mbox{]}~,
&&
\bar{\delta}_{G/T}\bar{\psi}
=\frac{1}{2}\mbox{[}\bar{\psi},\bar{\psi} \mbox{]}~~.
\label{eq:cosetBRST}
\eeqa
Contributions other than the above described  vanish.
Note that the above BRST-like symmetry is that obtained by
twisting the N=2 superconformal symmetry
realized by the Kazama-Suzuki formalism \cite{KazamaSuzuki}.
We will call it $G/T$-BRST symmetry.

{}~

             The path-integral quantization of the model (\ref{eq:action1}) is
performed along the formalism developed in \cite{GK},\cite{KS},
by utilizing the concept of
"chiral gauge transformation",
that is, complex gauge transformation defined for
any $T^{c}$-valued function.
Under the chiral gauge transformation,
the bosonic part of the action (\ref{eq:action1})
satisfies the "Polyakov-Wiegmann identity";
\beqa
S_{G}(^{\Omega}g,{}^{\Omega}\!A)
=S_{G}(g,A)-S_{G}(\Omega^{\dag}\Omega,A)~~~,
\eeqa
where $\Omega$ is $T^{c}=(U(1)^{c})^{\otimes (n-1)}$-valued, and
\beqa
&&^{\Omega}g=\Omega
^{\dag -1}g \Omega^{-1}\n
&&(^{\Omega}A)_{z}
=-\partial_{z} \Omega \Omega^{-1}+ A_{z}~~~.
\label{eq:Ttrans}
\eeqa
The "Polyakov-Wiegmann identity" tells us that
the bosonic part (\ref{eq:action2}) has the gauge invariance for the
vectorial direction, that is,
the $T=U(1)^{\otimes (n-1)}$ gauge transformation.
But when $\Omega$ is $T^{c}/T$-valued ($^{i.e}$ the axial direction),
The theory suffers the chiral anomaly at the classical level.
Of course the fermionic part (\ref{eq:action3}) has no anomaly
at the classical level. But, at the quantum level, their determinant
behaves anomalously for the axial direction
of the chiral gauge transformation.

                  Now suppose we parametrize the gauge field $A$ as
\beqa
A_{z}=-\partial_{z} \Omega \Omega^{-1},
\label{eq:gauge}
\eeqa
where $\Omega=e^Xe^{iY}$ is $T^{c}=(U(1)^{c})^{\otimes (n-1)}$-valued
($e^X, e^{iY}$ are $T^{c}/T, T$-valued respectively).
With this parametrization (\ref{eq:gauge})
we can perform the path-integration of the model.
Note that,
while $\int DY$,
being simply the gauge volume,
can be dropped without any trouble,
we cannot drop
$\int DX$
because of the chiral anomaly.
After some calculations of the anomalies
we obtain the following relation \cite{SN},\cite{SN2};
\beqa
&&
\int \frac{DADgD(\chi,\bar{\chi},\psi,\bar{\psi})}{\mbox{gauge volume}}
e^{-I(g,\chi,\psi,A)} \n
&&~~~~~~~~~~~
=\int DXDgD(\chi,\bar{\chi},\psi,\bar{\psi})D(\zeta,\bar{\zeta},\xi,\bar{\xi})
e^{-I^{g.f}(g,X,\chi,\psi,\zeta,\xi)}  ,
\label{eq:pathintegral}
\eeqa
where the gauge fixed action $I^{g.f}$
is given by ;
\beqa
&&I^{g.f}
(g,\chi,\bar{\chi},\psi,\bar{\psi},\zeta,\bar{\zeta},\xi,\bar{\xi}) \n
&&~~~~~~=kS^{WZW}_{G}(g)
+S_{X}(X)+S_{\chi \psi}(\chi,\bar{\chi},\psi,\bar{\psi})
+S_{\zeta \xi}(\zeta,\bar{\zeta},\xi,\bar{\xi}) .
\label{eq:fixaction}
\eeqa
In the above
$X=(X_{1},\cdots,X_{n-1})$ are non-compact free bosons
originating  from the gauge field $A$
with the action\footnote{In what follows, we set $\rho$
the Weyl vector of $su(n)$ and define
$\alpha_{+}=\sqrt{k+n},~\alpha_{-}=-\frac{1}{\sqrt{k+n}}$.}
\beqa
S_{X}(X)=
\frac{1}{2 \pi i}\int_{\Sigma}d^2z~
\partial_{z}X \cdot \bar{\partial}_{\bar{z}} X
+\frac{1}{2 \pi}\int_{\Sigma} \sqrt{2} \alpha_{-} \rho\cdot X R~~.
\label{eq:actionX}
\eeqa
$\zeta=(\zeta_{1},\cdots,\zeta_{n-1}),\xi=(\xi_{1},\cdots,\xi_{n-1})$
are spin (1,0) ghost systems representating the Jacobian
of the parametrization (\ref{eq:gauge}).
We will call these (anti-)ghost fields
$\xi,\bar{\xi}~(\zeta_{z},\bar{\zeta}_{\bar{z}})$ as $T^{c}$-ghosts.
Their action is
\beqa
S_{\zeta \xi}(\zeta,\bar{\zeta},\xi,\bar{\xi})
=\frac{1}{2 \pi i}\int_{\Sigma}d^2z~
\zeta_{z} \cdot \bar{\partial}_{\bar{z}}\xi
+\bar{\zeta}_{\bar{z}}\cdot \partial_{z} \bar{\xi}~~~ .
\label{eq:actionTghost}
\eeqa

{}~

              At this stage the quantization of the model is straightforward,
since everything is expressed either in free fields, or
in terms of an ungauged $G$ WZW model.
First we note that, in order to extract the physical degrees of
freedom from our total system,
one needs to use the BRST symmetries of the combined system
(\ref{eq:fixaction}).
The gauge fixing for the chiral gauge transformation (\ref{eq:Ttrans})
should be characterized by the nilpotent BRST charge
$Q_{T^{c}}$ which generates $T^{c}$-BRST transformation;
\beqa
Q_{T^{c}}&=& \frac{1}{2 \pi i}
\int
G_{T^{c}}^{+}~~~, \n
G_{T^{c}}^{+}&=&
-\alpha_{-}(\xi \cdot H^{total}-\sqrt{2}\rho\cdot \partial \xi)~~~,
\label{eq:TBRST}
\eeqa
where $H^{total}$ is an anomaly free
$t$-current of the combined system ;
\beqa
H^{total}=\hat{H}+H_{X}~~~.
\label{eq:U(1)}
\eeqa
Here $\hat{H}$ is the current associated with the
$T=U(1)^{\otimes (n-1)}$ gauge symmetry,
which is composed of $H^{WZW}$, the Cartan current of the
$SU(n)$ WZW model and
$H_{\chi \psi}=\frac{1}{\sqrt{2}}
\sum_{\alpha \in \Delta_{+}}
\alpha \chi_{\alpha}\psi^{\alpha}$,
that of the coset ghosts
\footnote{They satisfy the OPEs:
$H^{WZW}_{i}(z)H^{WZW}_{j}(w)
\sim \frac{\frac{k}{2}\delta_{ij}}{(z-w)^2},~
H_{\chi \psi,i}(z)H_{\chi \psi,j}(w)
\sim \frac{\frac{n}{2}\delta_{ij}}{(z-w)^2}$.}.
$H_{X}=\alpha_{+}\partial X$
is the current of the gauge field
\footnote{$H_{X,i}(z)H_{X,j}(w)\sim
\frac{-\frac{k+n}{2}\delta_{ij}}{(z-w)^2}$.}.

                      As we described before, the nilpotent BRST charge
$Q_{G/T}$ which generates the $G/T$-BRST transformation (\ref{eq:cosetBRST})
can be constructed from one of the supersymmetry operators realized
by the Kazama-Suzuki formalism \cite{KazamaSuzuki};
\beqa
Q_{G/T}&=&\frac{1}{2 \pi i}
\int
G_{G/T}^{+}\n
G_{G/T}^{+}&=&
-\alpha_{-}\sum_{\alpha \in \Delta_{+}}
\psi^{\alpha}
(J^{WZW}_{\alpha}-
\sum_{\beta \in \Delta_{+},\alpha < \beta}
\psi^{\beta}\chi_{\alpha +\beta})~~~.
\label{eq:KSsuper}
\eeqa
Note that the BRST currents
$G_{T^{c}}^{+}$ (\ref{eq:TBRST}) and $G_{G/T}^{+}$ (\ref{eq:KSsuper})
anticommute each other.
Then the coresponding BRST charges
$Q_{T^{c}},Q_{G/T}$ anticommute.
The physical observables can be characterized as the cohomology classes
of the double complex, whose differential is given by the total
BRST charge $Q^{total}=Q_{T^{c}}+Q_{G/T}$.
In what follows, assuming the degeneration of the corresponding
spectral sequence, we will study this cohomological problem
by taking cohomology first by $Q_{G/T}$ and then by $Q_{T^{c}}$.

                     The total energy-momentum (EM) tensor of the combined
system (\ref{eq:fixaction}) is easily derived;
\beqa
T^{tot}=T_{WZW}+T_{\chi \psi}+T_{X}+T_{\zeta \xi} ~~,
\eeqa
where $T_{WZW}$ is the EM tensor of the $SU(n)$ WZW model,
and $T_{X},T_{\chi \psi},T_{\zeta \xi}$ are those of
the gauge field, the coset ghosts, the $T^{c}$-ghosts respectively.
Their explicit forms are given by ;
\beqa
&&T_{X}=-\partial X \cdot \partial X
+\sqrt{2}\alpha_{-}
\rho\cdot \partial^2 X~~~,\n
&&T_{\chi \psi}=-\sum_{\alpha \in \Delta_{+}}
\chi_{\alpha}\partial \psi^{\alpha}~,~~~~
T_{\zeta \xi}=-\zeta_{z}\cdot \partial \xi~~~.
\eeqa
{}From the above explicit forms we can see that $T^{tot}$ has vanishing
central charge.
So the model is a TFT.
Moreover $T^{tot}$ can be written as the following BRST exact
form;
\beqa
T^{tot}
=\{ Q^{tot}~,~G_{G/T}^{-}+G_{T^{c}}^{-} \}~~~,
\label{eq:exactT}
\eeqa
where $G^{-}_{G/T}$ is the spin 2 fermionic current
constructed from the other supersymmetry operator
realized by the Kazama-Suzuki formalism,
and $G_{T^{c}}$ is also a spin 2 fermionic current.
Its explicit form is;
\beqa
G^{-}_{T^{c}}=-\alpha_{-}
(\zeta_{z} \cdot (\hat{H}-H_{X})
-\sqrt{2}\rho\cdot \partial \zeta_{z})~~.
\eeqa
We observe that the BRST currents
$G_{G/T}^{+},G_{T^{c}}^{+}$
anti-commute with $G^{-}_{T^{c}},G^{-}_{G/T}$
respectively.\footnote{$G_{G/T}^{+}(z)G_{T^{c}}^{-}(w)\sim
G_{T^{c}}^{+}(z)G_{G/T}^{-}(w)\sim \mbox{regular at $z=w$}$.}
Hence we can factorize the BRST-exact form (\ref{eq:exactT})
into two commuting pieces;
\beqa
T^{tot}=T_{G/T}+T_{T^{c}}~~,
\label{eq:factori}
\eeqa
where we set
\beqa
T_{G/T}&=&\{ Q_{G/T}~,~G_{G/T}^{-}\}~\n
&=& T_{WZW}+T_{\chi \psi}
-\frac{1}{k+n}\hat{H}\cdot \hat{H}
+\frac{\sqrt{2}}{k+n}\rho \cdot \partial \hat{H}~~,\n
T_{T^{c}}&=&\{ Q_{T^{c}}~,~G_{T^{c}}^{-}\}~~\n
&=& T_{X}+T_{\xi \zeta}
+\frac{1}{k+n}\hat{H}\cdot \hat{H}
-\frac{\sqrt{2}}{k+n}\rho \cdot \partial \hat{H}~~.
\eeqa
Note that
$\{ T_{G/T},G_{G/T}^{\pm},J_{KS}\}$
generate the topological conformal algebra (TCA)
obtained by twisting the Kazama-Suzuki model for the flag manifold
$G/T=SU(n)/U(1)^{\otimes (n-1)}$
(central charge
$C_{KS}=\frac{3}{2}n(n-1)-\frac{12 \rho^2}{k+n}$),
with the definition of the $U(1)$ current,
\beqa
J_{KS}=\frac{2\sqrt{2}}{k+n}\rho\cdot \hat{H}+N_{\chi \psi}~~,
\label{eq:KSU(1)}
\eeqa
where $N_{\chi \psi}$ is the ghost number current for the coset ghosts
$\chi_{z},\psi$.

                     Suppose we firstly take cohomology by $Q_{G/T}$.
Then the reduced system (with respect to $Q_{G/T}$) will
be characterized by the EM tensor $T_{T^{c}}$.
Observing that $T_{T^{c}}$ has vanishing central charge and
can be written as the BRST-exact form
$T_{T^{c}}=\{ Q_{T^{c}}~,~G_{T^{c}}^{-}\}$,
we can ask if it is possible to characterize this
reduced system by an appropriate TCA.
There exists such a TCA.
Namely we can see that
$\{ T_{T^{c}},G_{T^{c}}^{\pm},J_{T^{c}}\}$
generate a TCA with a $U(1)$ current;
\beqa
J_{T^{c}}=-\frac{2\sqrt{2}}{k+n}\rho\cdot H_{X}+N_{\zeta \xi}~~,
\label{eq:CGU(1)}
\eeqa
where $N_{\zeta \xi}$ is the ghost number current for the $T^{c}$-ghosts
$\zeta, \xi$.
{}From the OPE among this $U(1)$ current, we see the above TCA
is obtained by twisting a N=2 superconformal algebra with central
charge
$C_{T^{c}}=3(n-1)-\frac{12 \rho^2}{k+n}$.

{}~

                                Let us now connect our analysis
to those using path-integral formulation.
It has been pointed out in \cite{Witten}
that the path-integration in BRST invariant theories is actually
reduced to the integral over the fixed point sets under
BRST transformation.
We conjecture that,
in our case when the BRST operator is identified as
$Q_{G/T}$ (\ref{eq:KSsuper}),
the second piece in (\ref{eq:factori}) describes
the theory reduced on to the BRST fixed points.

                         In order to discuss this reduction,
it is convenient to utilize the free field realization of the
$SU(n)$ WZW model ($c.f$ \cite{IK}).
Introducing compact free bosons
$\phi =(\phi_1,\cdots,\phi_{n-1})$
and bosonic
$(\beta_{\alpha},\gamma^{\alpha})$
systems $(\alpha \in \Delta_{+})$,
the $SU(n)$ WZW model may be described by the action;
\beqa
kS_{G}^{WZW}
(\phi,\beta,\bar{\beta},\gamma,\bar{\gamma})
=S_{\phi}(\phi)
+S_{\beta \gamma}(\beta,\bar{\beta},\gamma,\bar{\gamma}) ~~~,
\eeqa
where
\beqa
&&S_{\phi}(\phi)=\frac{1}{2 \pi i}\int_{\Sigma}d^2z
\partial_{z}\phi \cdot \bar{\partial}_{\bar{z}} \phi
+\frac{i}{2 \pi}\int_{\Sigma} \sqrt{2} \alpha_{-}
\rho\cdot \phi R ~~,\n
&&S_{\beta \gamma}(\beta,\bar{\beta},\gamma,\bar{\gamma})
= \frac{-1}{2 \pi i}\int_{\Sigma}d^2z
\sum_{\alpha \in \Delta_{+}}
(\beta_{z,\alpha}\bar{\partial}\gamma^{\alpha}
+\bar{\beta}_{\bar{z}}^{\alpha}\partial \bar{\gamma}_{\alpha})~~~.
\eeqa
Note that the BRST current $G^{+}_{G/T}$ (\ref{eq:KSsuper})
is essentially the same as that appears
in the study of the quantum Hamiltonian
reduction $\grave{a} la$ Drinfeld-Sokolov \cite{BO},\cite{FF}.
We can prove that the bosonic ghosts
$(\beta_{\alpha},\gamma^{\alpha})_{\alpha \in \Delta_{+}}$
and the coset ghosts
$(\chi_{\alpha},\psi^{\alpha})_{\alpha \in \Delta_{+}}$
form a pair of the $G/T$-BRST doublets
($c.f$ \cite{BO} \cite{FF}), and so these fields decouple
from the theory (the "quartet confinements").
Especially the following relation holds;
\beqa
S_{\beta \gamma}(\beta,\bar{\beta},\gamma,\bar{\gamma})
+S_{\chi \psi}(\chi,\bar{\chi},\psi,\bar{\psi})
= \{ Q_{G/T}+\bar{Q}_{G/T}~,~V \}~~~,
\label{eq:confine}
\eeqa
where $V$ is an appropriate functional.
Strictly speaking we have checked the above relation
only in the case of $n=2$.
But, taking account of the results obtained in \cite{FF},
it should hold in general.

                       The fixed points of the $G/T$-BRST transformation
(\ref{eq:cosetBRST}) is easily derived because the current
$J^{WZW}$ which appears in the definition of the BRST current
$G^{+}_{G/T}$ (\ref{eq:KSsuper})
can be realized in terms of the bosonic ghosts
$(\beta_{\alpha},\gamma^{\alpha})_{\alpha \in \Delta_{+}}$.
Taking also the anti-holomorphic part into account,
these fixed points may be characterized by;
\beqa
\beta_{z, \alpha}=\gamma^{\alpha}=0~~,~
\bar{\beta}_{\bar{z}}^{\alpha}=\bar{\gamma}_{\alpha}=0~~~,
\label{eq:fixpoint}
\eeqa
for any $\alpha \in \Delta_{+}$. This is consistent
with  the quartet confinements (\ref{eq:confine}).

                            On the fixed points (\ref{eq:fixpoint})
the residual dynamical degrees of freedom are
$\phi,X$ and the $T^{c}$-ghosts $\zeta,\xi$.
The action of these fields may be derived from the gauge fixed action
$I^{g.f}$ (\ref{eq:fixaction})
by dropping out the irrelevant terms (\ref{eq:confine}).
Introducing a complex boson
$\varphi =\phi +iX $, it is given by;
\beqa
&&I_{fixed~point}(\varphi,\varphi^{*},\zeta,\bar{\zeta},\xi,\bar{\xi}) \n
&&~~~~~=
\frac{1}{4 \pi i}\int_{\Sigma}d^2z~
\partial_{z}\varphi \cdot \bar{\partial}_{\bar{z}} \varphi^{*}
+\partial_{z}\varphi^{*} \cdot \bar{\partial}_{\bar{z}} \varphi
+\frac{i}{2 \pi}\int_{\Sigma} \sqrt{2} \alpha_{-}
\rho\cdot \varphi^{*} R  \n
&&~~~~~~~~+\frac{1}{2 \pi i}\int_{\Sigma}d^2z~
\zeta_{z} \cdot \bar{\partial}_{\bar{z}}\xi
+\bar{\zeta}_{\bar{z}}\cdot \partial_{z} \bar{\xi} .
\label{eq:fixptaction}
\eeqa
The action $I_{fixed~point}$
is that of the Coulomb gas realization of the topological
minimal model in the case of $n=2$,
and its generalization for $n \geq 3$.
In fact the system (\ref{eq:fixptaction})
is characterized by the following TCA;
\beqa
G^{+}_{C.G}&=&
\sqrt{2}\alpha_{-}\rho\cdot \partial \xi
+i\xi \cdot \partial \varphi^{*}~~, \n
G^{-}_{C.G}&=&
\sqrt{2}\alpha_{-}\rho\cdot \partial \zeta_{z}
+i\zeta_{z}\cdot \partial \varphi ~~, \n
T_{C.G}&=&
-\partial \varphi \cdot \partial \varphi^{*}
+i\sqrt{2}\alpha_{-}\rho\cdot \partial^{2}\varphi^{*}
-\zeta_{z}\cdot \partial \xi ~~,\n
J_{C.G}&=&-i\sqrt{2}\alpha_{-}\rho \cdot \partial
(\varphi-\varphi^{*})+N_{\zeta \xi}~~.
\label{eq:CGTCA}
\eeqa
{\em Modulo $G/T$-BRST exact terms this TCA is equal to
that generated by $\{T_{T^{c}},G_{T^{c}}^{\pm},J_{T^{c}}\}$
which describes the reduced system in the operator formulation.}
\footnote{Because $H^{total},\hat{H}$ (\ref{eq:U(1)}) are
equal to $i\alpha_{+}\partial \varphi^{*},i\alpha_{+}\partial \phi$
respectively modulo $G/T$-BRST exact terms.}
{\em Thus the TFT (\ref{eq:fixptaction})
is preceisely what we wish to have.}

{}~

                             The TFT on the $G/T$-BRST fixed points
(\ref{eq:fixptaction}) will play an important role
in the study of the "WG-gravity" \cite{WG}.
These will be studied in separate papers \cite{SN},\cite{SN2}.
Here we give some comments:
In the case of $n=2$
the TFT on the fixed points is precisely the topological
matter system in the K.Li's theory of 2 dimensional gravity \cite{Li}.
And this TFT is equivalent to "the abelian model" \cite{Witten},
that is, the topological $U(1)$ gauge theory coupled to
2 dimensional gravity.
Since our main purpose in this article is to
construct the TFT on the BRST fixed points,
we neglect its coupling to 2 dimensional gravity.
It is worth mentioning that
the process of twisting $N=2$ symmetry in 2 dimension
amounts to a mixing of Lorentz and internal
$U(1)$ rotations and redefinition of the quantum number
of the fields.
This indicates \cite{SN} the possibility that
the coupling to 2 dimensional topological gravity
can be performed via the back-ground gauge field
which we set trivial in (\ref{eq:gauge}).

                      Our discussion given in this article would carry
over to the case of $G=SL(n,R)$.
The $G/T$-BRST fixed points will be characterized by the same condition
as (\ref{eq:fixpoint}), and the TFT on these fixed points
is that described by the action (\ref{eq:fixptaction})
(with an appropriate change of $\alpha_{\pm}$).
It is known \cite{Witten2}
that the $SL(2,R)/U(1)$ gauged WZW model (with a slightly
different $U(1)$ embedding) describes a 2 dimensional
target space geometry with a metric\footnote{We follow
the notation in \cite{Witten2}.}
\beqa
ds^2=\frac{dudv}{1-uv}~~~,
\eeqa
which gives a 2 dimensional black hole.
The curvature of the metric blows up at $uv=1$.
Thus the space-time singularities appear at $uv=1$.
This region of the singularity is the $SL(2,R)/U(1)$-BRST
fixed points (\ref{eq:fixpoint})\footnote{To be precise,
the singularities and the fixed points are related by "duality"
\cite{Giveon}.}.
Hence, as discussed in \cite{Eguchi}, the TFT on the fixed points
will describe the region of the singularities in the black hole geometry.

\newpage
\section*{Acknowledgements}

                ~We would like to thank Prof.T.Eguchi for
several discussions and continuous encouragement during
this work was done.

{}~

\end{document}